\documentclass{sigchi}

\toappear{\scriptsize Permission to make digital or hard copies of all or part of this work for personal or classroom use is granted without fee provided that copies are not made or distributed for profit or commercial advantage and that copies bear this notice and the full citation on the first page. Copyrights for components of this work owned by others than ACM must be honored. Abstracting with credit is permitted. To copy otherwise, or republish, to post on servers or to redistribute to lists, requires prior specific permission and/or a fee. Request permissions from Permissions@acm.org. \\
\emph{CHI 2017, May 6-11, 2017, Denver, CO, USA.} \\
Copyright \copyright~2017 ACM ISBN 978-1-4503-4655-9/17/05\ ...\$15.00. \\
http://dx.doi.org/10.1145/3025453.3025853}

\usepackage{balance}  
\usepackage{graphics} 
\usepackage[T1]{fontenc}
\usepackage{txfonts}
\usepackage{mathptmx}
\usepackage[pdftex]{hyperref}
\usepackage{color, colortbl}
\usepackage{booktabs}
\usepackage{textcomp}
\usepackage{microtype} 
\usepackage{ccicons}  
\usepackage{parskip}
\usepackage{tabu}

\usepackage{todonotes}

\usepackage{float}

\def\plaintitle{Community-Empowered Air Quality Monitoring System}
\def\plainauthor{Yen-Chia Hsu, Paul Dille, Jennifer Cross, Beatrice Dias, Randy Sargent, Illah Nourbakhsh}

\def\plainkeywords{Citizen science; sustainable HCI; adversarial design; participatory design; community engagement; data visualization; air quality}

\makeatletter
\def\url@leostyle{%
  \@ifundefined{selectfont}{
    \def\UrlFont{\sf}
  }{
    \def\UrlFont{\small\bf\ttfamily}
  }}
\makeatother
\def\pprw{8.5in}
\def\pprh{11in}

\definecolor{linkColor}{RGB}{6,125,233}
\hypersetup{%
  pdftitle={\plaintitle},
  pdfauthor={\plainauthor},
  pdfkeywords={\plainkeywords},
  bookmarksnumbered,
  pdfstartview={FitH},
  colorlinks,
  citecolor=black,
  filecolor=black,
  linkcolor=black,
  urlcolor=linkColor,
  breaklinks=true,
}

\newcommand{\mypm}{\mathbin{\smash{%
			\raisebox{0.35ex}{%
				$\underset{\raisebox{0.5ex}{$\smash -$}}{\smash+}$%
			}%
		}%
	}%
}

\newcommand{\highlight}[1]{\textbf{#1}}
\newcommand{\highlightT}[1]{\textbf{#1}}

\definecolor{gray}{rgb}{0.85,0.85,0.85}
\definecolor{red}{rgb}{1,0,0}
\definecolor{blue}{rgb}{0,0,1}

\begin{document}

\title{\plaintitle}

\numberofauthors{1}
\author{%
  \alignauthor{\plainauthor\\
    \affaddr{The Robotics Institute, Carnegie Mellon University, Pittsburgh, U.S.A.}\\
    \email{$\lbrace$yenchiah, pdille, jcross1, mdias, rsargent, illah$\rbrace$@andrew.cmu.edu}}\\
}


\maketitle

\begin{abstract}
	
Developing information technology to democratize scientific knowledge and support citizen empowerment is a challenging task. In our case, a local community suffered from air pollution caused by industrial activity. The residents lacked the technological fluency to gather and curate diverse scientific data to advocate for regulatory change. We collaborated with the community in developing an air quality monitoring system which integrated heterogeneous data over a large spatial and temporal scale. The system afforded strong scientific evidence by using animated smoke images, air quality data, crowdsourced smell reports, and wind data. In our evaluation, we report patterns of sharing smoke images among stakeholders. Our survey study shows that the scientific knowledge provided by the system encourages agonistic discussions with regulators, empowers the community to support policy making, and rebalances the power relationship between stakeholders.

\end{abstract}

\category{H.5.m.}{Information Interfaces and Presentation (e.g. HCI)}{Miscellaneous}


\section{Introduction}

\begin{figure}[t]
	\centering
	\includegraphics[width=1\columnwidth]{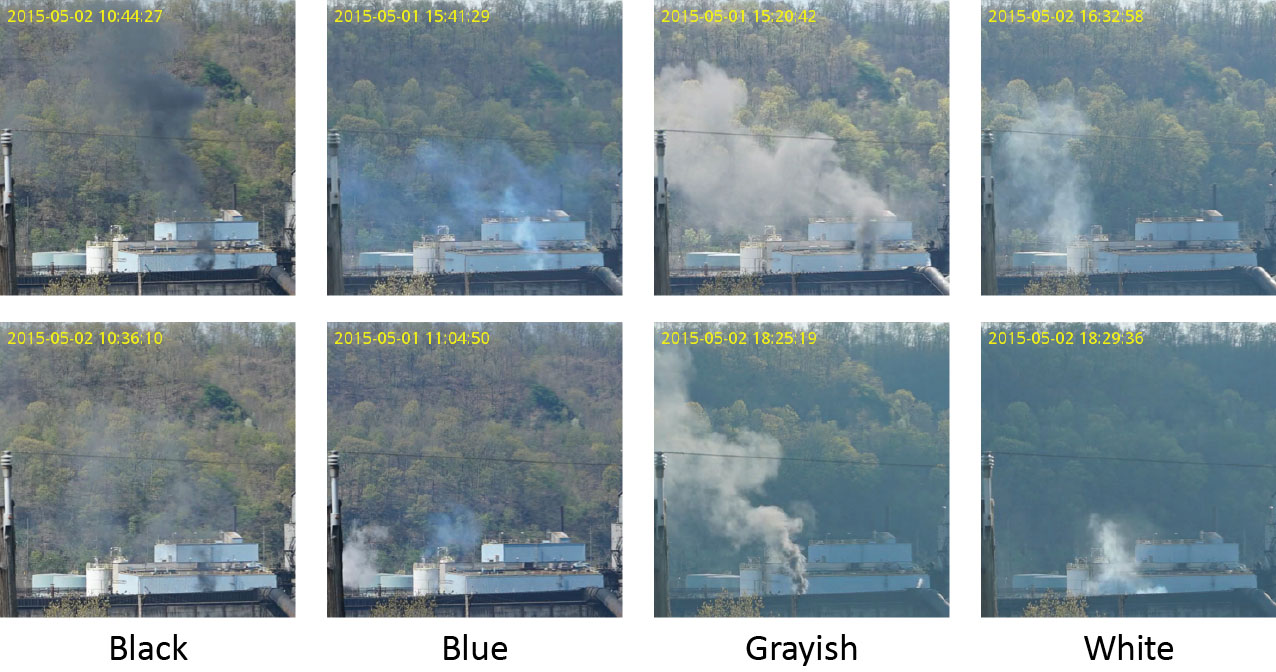}
	\caption{This figure shows different types of fugitive emissions.}
	\label{fig:smoke-type}
\end{figure}

\begin{figure*}[t]
	\centering
	\includegraphics[width=2.1\columnwidth]{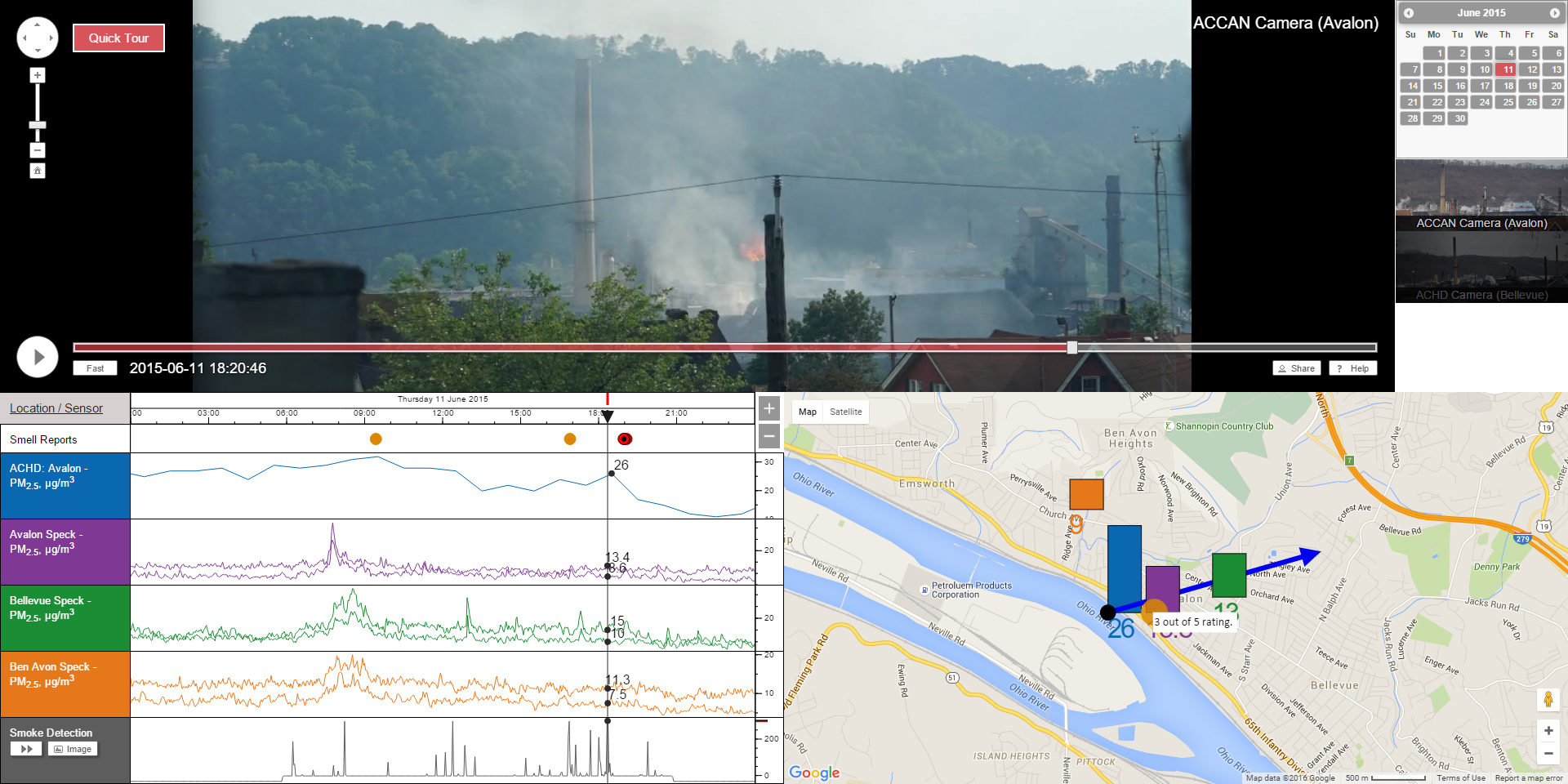}
	\caption{The user interface of the web-based air quality monitoring system. The top-left part is a zoomable and pannable viewer which shows the timelapse video. The bottom-left charts visualize crowdsourced smell reports, PM2.5 sensor readings, and automatic smoke detection results. The blue line shows readings from the sensor operated by the local health department. The purple, green, and orange lines shows readings from six sensors that we deployed in the community. The bottom-right map indicates wind speed (length of the blue arrow), wind direction (orientation of the blue arrow), and sensor locations (bar charts). The colors and heights of bar charts on the map correspond to the colors and readings on the line charts respectively.}
	\label{fig:UI}
\end{figure*}

Air pollution is a critical environmental issue for people who live near industrial sites. To address this problem, it takes communities a great effort to gather scientific evidence at a large spatial and temporal scale, which requires the assistance of information technology in collecting, curating, and visualizing various types of data. In our case, 70,000 residents near Pittsburgh suffer from air pollution caused by a coke (fuel) plant. Under some unusual situations, the coke plant leaks hazardous smoke irregularly, known as fugitive emissions (see Figure \ref{fig:smoke-type}), into the atmosphere. The resulting toxic emissions with fine particulates pose risks to health and have negative impacts to living quality \cite{Kampa2008, Pope2006}.

To address air pollution, residents formed the ACCAN (Allegheny County Clean Air Now) group. In several community meetings, residents mentioned that adults and children developed respiratory problems because of exposure to coke oven gas. In addition, residents must close windows at night because of irritating burning smells. They also said that the air quality was so poor that they could not exercise outside. To pursue environmental justice, the community took a series of actions, such as gathering evidence of violations and filing petitions to the government. They envisioned that these actions could raise public awareness about air quality issues and pressure the government to deal with air pollution problems.

To advocate for themselves in improving the local air quality, the community needed to gather convincing evidence in communicating with stakeholders. Traditionally, the community collected scientific data manually, which was time-consuming, error-prone, and offered limited scientific validity. The community lacked technological fluency and required the assistance of experts in setting up an automatic system to collect and archive data from various sources. Starting in January 2015, we aided the community to set up outdoor air quality sensors and live cameras pointed at the coke oven where smoke usually occurred. We also created an electronic process for capturing smell reports. To visualize hybrid data (sensor readings, smell reports, real-time high resolution imagery, and wind information), we developed a web-based air quality monitoring system. Community members could use the system to manually search for smoke in timelapse videos and use a thumbnail generator to create animated images. But searching and documenting all smoke emissions required manpower and took an impractical investment of time. Therefore, we implemented a computer vision tool to detect smoke and produce corresponding animated images (see Figure~\ref{fig:smoke-images}), which could then be curated in online documents and shared on social media. With the monitoring system, community members could tell stories with concrete scientific evidence about what happened (using animated smoke images) and how these events affected the local neighborhood (using sensor readings, smell reports, and wind information).

To evaluate community engagement, we analyzed the server logs, which store HTTP requests of thumbnails from August 2015 to July 2016. In addition, we conducted a survey study with the research question: does interacting with the air quality monitoring system increase community engagement in addressing air pollution concerns? We anticipated that the intervention of the system increases awareness, self-efficacy \cite{Bandura1977,Carroll2005}, and sense of community \cite{McMillan1986}, which are the three dependent variables in our survey study. Awareness means participants know a problem exists and has impact on daily lives. Self-efficacy means the strength of participants' belief in their ability to successfully reach the community's goal. Sense of community means participants feel they have influence in the community and a sense of belonging. We form three corresponding hypotheses: interacting with the system improves the ability to perceive air quality problems, strengthens the belief that the ACCAN community can reach its goal of improving air quality, and makes people think that they are influential and fit in the community. The independent variables are involvement, age range, and education level. Involvement is the level of participation, such as exploring, documenting, and sharing data from the system. 

In this paper, we explore the formation and use of scientific knowledge in citizen empowerment via the intervention of information technology. Our design principle is to stimulate critical discussions and confront the current unbalanced power relation between stakeholders. We begin by explaining the research scope and reviewing similar projects. Then, we describe the design process and the implemented web-based air quality monitoring system. In addition, we discuss the results of smoke image usage from server logs and survey study. Finally, we provide insights in developing systems to empower data-driven community action and conclude with limitations. Our contributions are:
\begin{itemize}
	\item Detailed documentation of a worked example which used scientific data from heterogeneous sources to critically reveal, question, and challenge environmental conditions.
	\item Analysis of community behavior changes after the intervention of information technology and participatory design.
	\item Analysis of how the community uses smoke images over a long-term participation period (12 months).
	\item Insights for researchers to develop environmental monitoring systems that combine politics, community, and information technology.
\end{itemize}

\section{Related Work}

Citizen science \cite{Irwin1995,DickinsonBonney2012} and sustainable human-computer interaction \cite{DiSalvo2009,DiSalvoSengers2010,Brynjarsdottir2012,Blevis2007,Mankoff2007} are growing trends in addressing community concerns around local specific problems, such as air pollution. The core idea is to empower amateurs and professionals to produce scientific knowledge through public engagement \cite{EU2013,McKinley2015}, which is different from conventional communication methods, such as newsletters or public hearings. Scientific knowledge can contribute to the well-being of communities and has two major values: science education \cite{DickinsonBonney2012,MillerRushing2012,Silvertown2009,BonneyCooper2009,DickinsonShirk2012}, which increases public understanding of science by spreading knowledge among common people, and participatory democracy \cite{Irwin1995,Greaves1980,Wilsdon2005,Stilgoe2009,Irwin2001}, which promotes the idea that citizens can participate in policy-making using scientific evidence. Our work focuses on the latter.

Citizen science has various participation levels \cite{BonneyBallard2009,Wiggins2011,Haklay2013,Shirk2012}. They include defining problems, developing plans, providing data, analyzing data, and making decisions. Consider participation along a spectrum from citizens as tools to citizens as scientists. At one end, scientists treat volunteers as tools that provide or interpret data, which is also called crowdsourcing. One typical example is Galaxy Zoo \cite{Lintott2008}, where participants classify a large amount of galaxies online. At the other end, scientists treat volunteers as collaborators over the entire project life cycle. Our work takes the concept of citizens as scientists, where volunteers and scientists establish a strong partnership through collaboration and engagement.

Citizen science can be led by a central organization, multiple stakeholders, or a community. These correspond to three governance structures respectively: top-down, multi-party, and bottom-up \cite{Conrad2011}. One example of the top-down approach is The Neighborhood Networks Project \cite{DiSalvo2008, DiSalvo-Louw-2009}, a participatory design practice which uses sensing technology to engage residents in collecting environmental data and making critical discussions about local environmental issues. The project is led by researchers from an academic institution with pre-designed public engagement procedures. In contrast, we collaborated with the community via the bottom-up approach, where communities themselves initiate, organize, and lead grassroots movements about local specific issues. An example of the bottom-up approach is the Bucket Brigade project \cite{GCM-2016}, which provides a low-cost device for citizens to measure their local air quality. During our collaboration with the community, besides applying the concept of citizens as scientists, which asks what common people can do for professional scientists, we emphasize the importance of scientists as citizens \cite{Irwin1995,Wilsdon2005,Stilgoe2009,Haklay2013}, which asks what professional scientists can do for common people. Our work explores how scientists can engage in social and ethical issues that are promulgated by citizens.

Modern citizen science projects gain scientific knowledge at a large spatial and temporal scale. This requires the investment of information technology to collect, curate, and visualize data from heterogeneous sources, such as video, sensors, and crowds \cite{BonneyCooper2009,Newman2012}. Developing interactive computational tools to improve technological fluency among citizen scientists and to foster communication for participatory democracy is an ongoing challenge \cite{Bonney2014,McKinley2015}. This research focuses on empowering local communities in pursuing environmental justice collaboratively through the intervention of information technology, which falls into the field of designing computational tools to support democratizing scientific knowledge. The largest limitation in prior work is that they often focus on either video, sensor, or crowdsourced data. However, in our context, communities need to interpret the cause and effect of emissions from an identified pollution source. This requires collecting, curating, visualizing, comparing, and making sense of hybrid scientific data, which include live imagery, sensor readings, image recognition results, smell reports, wind direction, and wind speed. We are unaware of the existence of systems that can tackle the complexity of data required for this task.

\subsection{Understanding Video Data Using Crowdsourcing}
There are tools which focus on understanding video data using crowdsourcing. SynTag \cite{Hsu2012} is a real-time collaborative tagging system for labeling presentation videos. Audiences can tag "good", "question", or "disagree" during or after the presentation. These tags are shown on a line chart, which provides indices for video segments. Glance \cite{Lasecki2014} is a video coding tool which asks crowdworkers on Mechanical Turk to label small clips in parallel. Without searching the entire video, users can use the aggregated crowdsourcing labels to quickly identify events. Further extending this concept, Zensors \cite{Laput2015} is an image recognition tool using cameras on mobile devices, which combines crowdsourcing and computer vision to answer user defined questions. Initially, answers are provided by workers on Mechanical Turk. The tool then uses these answers as labels to train a computer vision classifier, which takes the image recognition task when its accuracy reaches a threshold for a certain period. These tools provide us with the concept that using other data sources to index archived videos can help users in making sense of video content efficiently.

\subsection{Gathering and Managing Sensor Data}
Other work focuses on gathering and managing sensor data. Kuznetsov et al. in 2011 \cite{Kuznetsov2011} developed a monitoring system which involves low-cost air quality sensors and web-based visualizations. In 2013, Kuznetsov et al. \cite{Kuznetsov-2013} designed a bio-electronic soil sensor to detect bacterial activities and visualized the data by using LED matrices and an LCD screen on a wooden enclosure. Kuznetsov et al. in 2014 \cite{Kuznetsov2014} presented a low-tech and low-cost paper sensor for collecting particulate pollution in the air. Kim et al. \cite{KimPaulos2013} described an indoor air quality monitoring system in domestic environments. Tian et al. \cite{Tian-2016} implemented a low-cost wearable sensor to measure airborne particles and a mobile application for visualizing the air quality data. These works emphasize studying the impact after deploying sensors on multiple communities by analyzing changes of user behaviors and ways that users interact with technology. User studies show that sensing technology with well-crafted visualizations can engage local communities to participate in political activism using concrete scientific evidence.

\subsection{Collecting and Curating Crowdsourced Data}
Several works focus on collecting and curating crowdsourced data. Creek Watch \cite{KimRobson2011} is a monitoring system which enables collecting water flow and trash data in creeks via cameras on mobile devices. EOL (Encyclopedia of Life) \cite{RotmanProcita2012} is a platform for curating biological content. Participants can comment and make "trust," "untrust," or "hide" decision on data. Content providers can then improve the data based on the collaborative feedback. Sensr \cite{KimMankoff2013} is a framework for creating data collection and management applications on mobile devices without programming skills. Project managers can use the tool to create a campaign website around a specific issue, such as air quality, and community members can report data, such as images, via a mobile application. eBird \cite{Sullivan2014, Sullivan2009} is a crowdsourcing tool for birdwatchers, scientists, and policy makers to collect, store, visualize, and analyze bird data. These works demonstrate that, beyond collecting data, it is important to understand what data stakeholders need, how to make data useful, and how to effectively deliver data. They consider tools as an integrated system, which supports different levels of participation, rather than individual and separate components.

\section{Design Process and Challenges} 

Our main goal is to use information and communication technology to democratize scientific knowledge by empowering citizens to collect and interpret data as evidence in taking political action. This falls in the context of design for democracy \cite{DiSalvo2010}, which has in general two opposing approaches: consensus and agonism. The design principle of consensus is to support structured deliberation. One project that uses the consensus approach is eBird \cite{Sullivan2014, Sullivan2009}, which is a tool for birdwatchers, scientists, and policy makers to collect, visualize, and analyze bird data collaboratively. Another example is the Community Resource Messenger \cite{LeDantec-2011}, which applies ubiquitous computing at a shelter for homeless mothers to facilitate the communication between staff and residents. In contrast, our design principle is adversarial design \cite{DiSalvo2012}, which is based on agonism \cite{Mouffe2000}. Adversarial design promotes critical political discussions and challenges the current unbalanced power structure between citizens, governments, and businesses. One project that uses adversarial design is Feral Robot \cite{Lane-2006}, which is a low-cost mobile sensor for grassroots communities to collect, map, and present chemical pollution data in a local park. Our design purpose is not to support the mechanism and procedure of governance, but to improve the condition of society. The information technology is not used to solve the environmental problem, but to provide technology affordance \cite{Gaver1991} for seeking and revealing the condition (form, function, economy, and time \cite{Pena2012}) of the problem.

We began by participating in monthly community meetings to understand the context of air pollution issues. The community was taking a series of actions, such as reporting industrial smells and filing petitions to the local health department and the EPA (Environmental Protection Agency). Our roles were as supporters, which use information technology to assist the citizen-led grassroots movement around local air quality issues, and as researchers, which study the effect of the technological intervention.

The successfulness of the intervention of information technology is highly dependent on community engagement \cite{Stevens2014}, the involvement of citizens in local neighborhoods. During initial discussions with the community, we found that the most significant gap in community engagement is the lack of scientific evidence. For instance, it was difficult for residents to report the exact time when an air quality violation occurred and its environmental impact to government regulators. Therefore, we proposed building an air quality monitoring system, which could afford exploring, archiving, presenting, and sharing scientific evidence among stakeholders.

The problem that the community dealt with is in nature wicked \cite{Rittel1973, Conklin-2005}. One characteristic of a wicked problem is that it cannot be fully observed, which means that solving a subset of a wicked problem reveals new ones. Based on this idea, we argue that our work requires an iterative design approach to handle and solve design challenges step by step. Thus, we adopt the community-based participatory design approach \cite{DiSalvo2013}. It is iterative in the sense that citizens and developers explore design options collaboratively.

We collaborated closely with the community and implemented system features based on iterative feedback from community members. There were two major design challenges in setting up the monitoring system. First, the community did not have sufficient technological fluency. Our system had to curate and visualize data in a way that users could easily perceive and document the seriousness of smoke emissions and their impacts to local neighborhoods. Second, this work had a timing issue, where residents had to form and use strong scientific evidence to convince regulators on a planned community meeting with the local health department and the EPA. These challenges served as constraints that affected our design decisions.

\section{System} 
	
We now explain system components together with three design iterations, which naturally emerged during the design process. The number of iterations depends on the complexity of the wicked problem \cite{Rittel1973, Conklin-2005} that the community tackles. Each iteration contained system features which were implemented based on the challenges revealed iteratively.

\subsection{First Iteration:}
\subsubsection{Interactive Web-based Timelapse Viewer}
\begin{figure}[t]
	\centering
	\includegraphics[width=1\columnwidth]{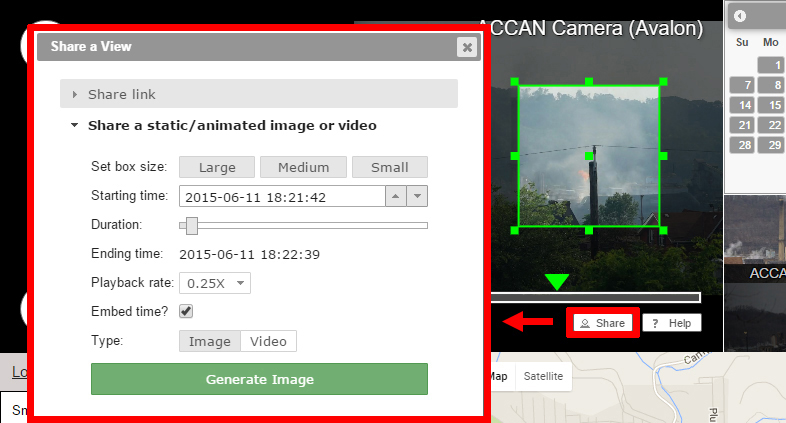}
	\caption{Clicking the share button on the timelapse viewer on the main user interface (see Figure \ref{fig:UI}) shows the thumbnail tool, which is used for generating sharable animated images. Users can edit the image size by resizing the green box on the viewer. The dialog window provides adjustable parameters, such as starting time and duration of the animated image.}
	\label{fig:thumbnail-tool}
\end{figure}

Starting in January 2015, we installed a live camera which was oriented towards the coke plant from a volunteer home. The live camera takes a high quality image every 5 seconds for a total of 17,000 each day. We streamed the time-series imagery to our servers and used an open source tool developed by Sargent et al.~\cite{Sargent2010} to process these images into multi-resolution video tiles. The tool was implemented in JavaScript/HTML and provided an \highlightT{interactive web-based timelapse viewer} (top-left part of Figure~\ref{fig:UI}) where users could search for fugitive emissions by panning, zooming, and playing the video. The viewer loaded and showed the video tile corresponding to the zoom and pan level. Users could share a particular view or use the thumbnail tool to generate sharable animated images (see Figure~\ref{fig:thumbnail-tool}). After we developed the web-based viewer, community members were excited and shared screenshots with each other via emails. At that time, the community pointed out two major challenges. Static images such as screenshots could not represent the dynamics and persistent time quality of smoke emissions. In addition, although smoke images indicated the source of air pollution, they did not show the impacts to local air quality. These challenges led to the next design iteration.

\subsection{Second Iteration:}
\subsubsection{Thumbnail Generator and Sensor Data Visualization}
To address the emergent challenges, we implemented a \highlightT{thumbnail generator}, which allowed community members to create and document animated smoke images as visual evidence. We also visualized $\mathrm{PM_{2.5}}$ (particle pollution) data from a sensor station operated by the local health department. In addition, we visualized smell reports which were collected via a Google Form, only available to community members. In the form, we asked community members to rate the severity of the pollution odors from 1 to 5, with 5 being the worst. The form was disseminated to the community via a Google Groups email and phone calls. The \highlightT{visualization of air quality data and smell reports} showed how smoke emissions affected the living quality of the community. With these new features, residents could compare smoke images together with sensor and crowdsourced data to identify correlations. We recorded a tutorial video and taught residents how to use these features during community meetings. The community was using the tool to find, generate, and share animated smoke images. However, searching smoke emissions manually from a large amount of time-series imagery was laborious and time-consuming. Moreover, the government-operated sensor station reported data only once per hour, which had difficulties in identifying air quality changes over a shorter time period. Furthermore, the lack of visualized wind data and sensor locations hindered the ability to determine how pollutants affected the air quality hyperlocally. These challenges again led to another design iteration.

\subsection{Third Iteration:}
\subsubsection{Citizen Sensors, Computer Vision Tool, and Map Visualization}
\begin{figure}[t]
	\centering
	\includegraphics[width=1\columnwidth]{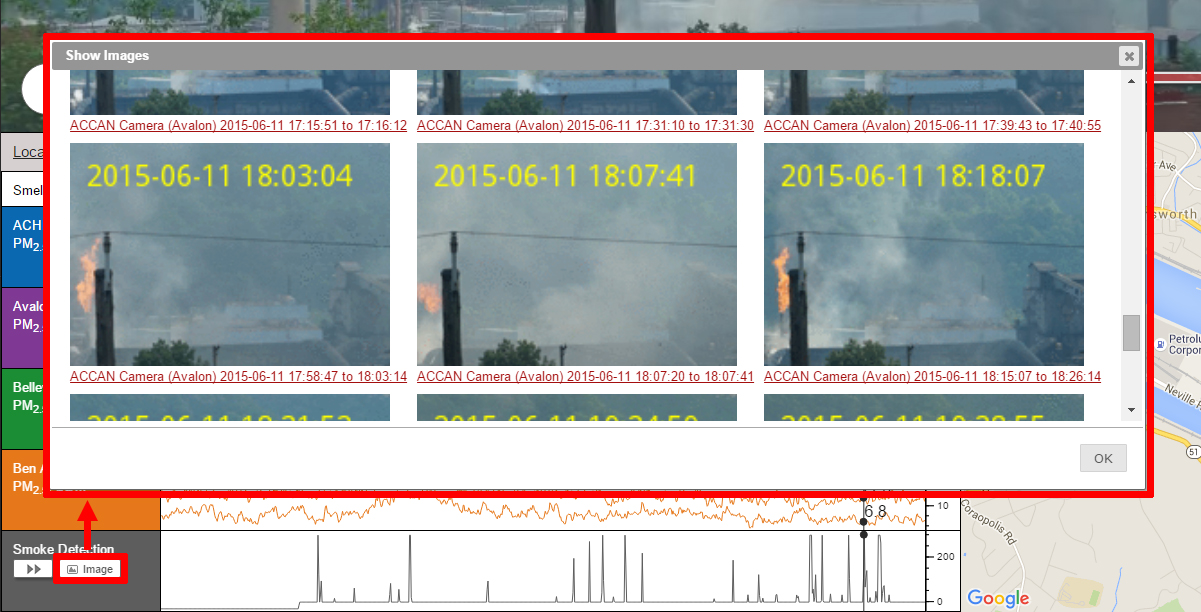}
	\caption{Clicking the image button on the line charts on the main user interface (see Figure \ref{fig:UI}) shows web links and animated images produced by the smoke detection algorithm. Users can quickly select representative images and insert them into an online document. Users can also click on a peak of a spike on the line chart to seek to a video frame with fugitive emissions.}
	\label{fig:smoke-images}
\end{figure}

To account for the challenges from the previous iteration, we deployed \highlightT{six commercial air quality sensors}~\cite{Speck,taylor2015} in local areas with finer time resolutions. These sensors reported $\mathrm{PM_{2.5}}$ data to our server via wireless Internet once per minute. The location of sensors and the Internet services were provided by community volunteers. Furthermore, we developed a \highlightT{computer vision tool} based on an existing smoke detection algorithm \cite{Hsu-2016} for finding fugitive emissions automatically. The algorithm identified the number of smoke pixels for each video frame at daytime (bottom chart in Figure~\ref{fig:smoke-images}) and automatically produced corresponding sharable animated images (see Figure~\ref{fig:smoke-images}). We also added a \highlightT{map visualization} for showing wind direction, wind strength, and sensor locations (bottom-right part of Figure~\ref{fig:UI}). All sensor data and smoke detection results were plotted on multiple charts (bottom-left part of Figure~\ref{fig:UI}). Users could use the charts as indicators for finding unusual events such as fugitive emissions. Clicking on a smell report or a peak of a spike on the chart jumped the video to the corresponding time. Users could also click on the image button near the smoke detection chart to bring up a dialog box with animated smoke images, which could be shared via social media or archived into a Google Doc.

The final design enabled community members to fully explore and compare data from heterogeneous sources (animated smoke images, finer air quality data, crowdsourced smell reports, and wind information). When residents noticed industrial smells like sulfur, they could use the timelapse viewer to check if the coke plant emitted smoke at a specific time. They could then compare sensor readings, smell reports, and wind data to verify if the emission came from the coke plant and affected the local air quality. With the system, the community could form and share convincing narratives grounded with scientific evidence aggregated from hybrid data.

\section{Evaluation}


Google Analytics evaluation of our website shows that from August 2015 to July 2016 there were 542 unique users, which contributed 1480 sessions. The average session duration was three minutes. We now discuss the image usage study for identifying how community members used animated images. Then we present the results of the survey study.

\subsection{Image Usage Study}

\begin{table}[b]
	\begin{center}
		\begin{tabu}{|l|c|}
			\hline \# of unique and viewed HG images & 135 \\
			\hline \# of views of all HG images & 477 \\
			\hline \# of unique and viewed AG images & 6745 \\
			\hline \# of views of all AG images & 11043 \\
			\hline \# of total views & 11520 \\
			\tabucline[2pt]{-}
			\hline \# of users who created HG images & 32 \\
			\hline \# of users who viewed HG images & 85 \\
			\hline \# of users who viewed AG images & 75 \\
			\hline \# of total users & 141 \\
			\hline
		\end{tabu} 
	\end{center}
	\caption{Summary statistics of animated smoke images and users. The "HG" and "AG" abbreviations mean "human-generated" and "algorithm-generated" respectively. The "\#" sign means "number of". We can see that the number of views of algorithm-generated images greatly exceeds the ones of human-generated images.}
	\label{tb:summary-image-user}
\end{table}

\begin{figure}[t]
	\centering
	\includegraphics[width=1\columnwidth]{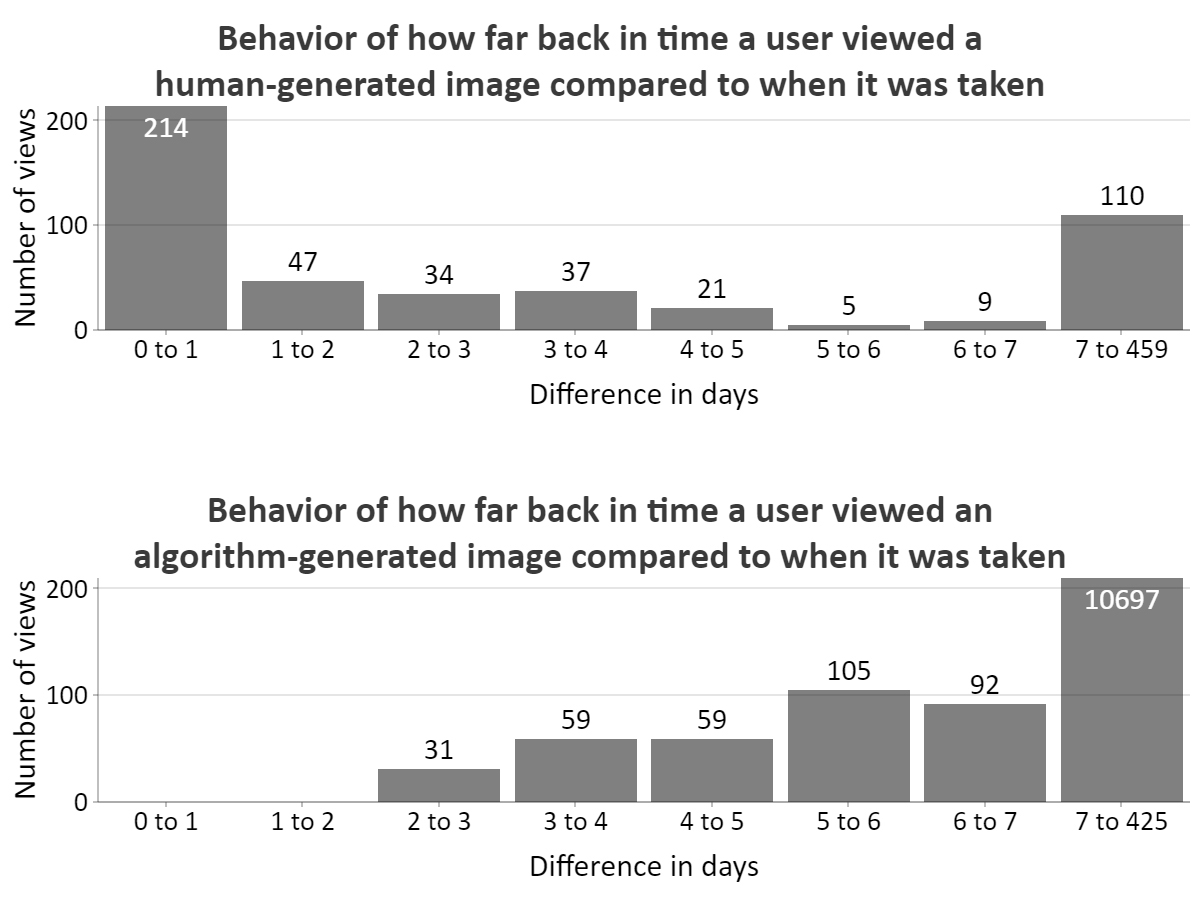}
	\caption{Behavior of how far back in time a user viewed a human-generated or algorithm-generated image compared to when it was taken. The x-axis is the difference in days (denote $D$) between the dates that an image was viewed and taken. Image views with small or large $D$ mean they are used for verifying if an event, such as fugitive emissions happened (e.g. fugitive emission) or reviewing previous events respectively. While human-generated images were often viewed in less than one day after events occur, algorithm-generated images were usually viewed at least a week after the events.}
	\label{fig:analysis-diff-time}
\end{figure}

\begin{figure}[t]
	\centering
	\includegraphics[width=1\columnwidth]{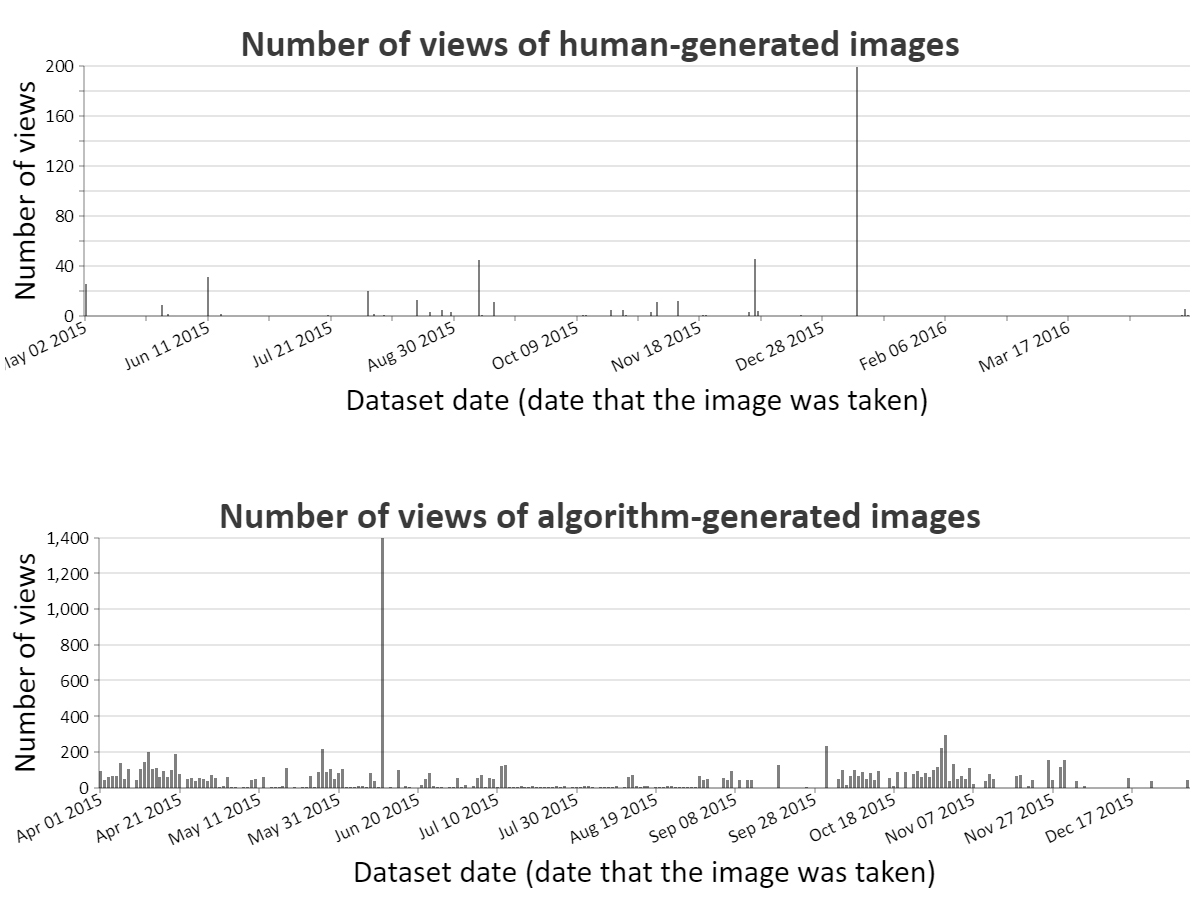}
	\caption{Number of views of human-generated or algorithm-generated images which are aggregated by dataset date. From these two graphs, we can see that the views of algorithm-generated images are more distributed across datasets, which means that users tend to use algorithm-generated images to explore events in different dates.}
	\label{fig:analysis-dataset-date}
\end{figure}

We evaluated the usage patterns of animated smoke images by parsing server logs. The logs stored HTTP requests of images from our server over an 11-month period from August 2015 to July 2016. Each request contained the source IP address, requested date, image URL, and browser type. Each image URL indicated its bounding box, size, time, and dataset. We first excluded all IP addresses from our research institute. Then for each HTTP request, we subtracted the requested date from the image taken date to get $D$, the difference in days, which indicated how far back in time a user viewed an image compared to when the image was taken. Table \ref{tb:summary-image-user} shows summary statistics of animated images and users. The number of views of algorithm-generated images greatly exceeds the ones of human-generated images. Next we discuss two sub-studies which focus on images and users.

\subsubsection{Image-based Sub-study}

For the image-based sub-study, we separated images into two sets: created by human or created by the computer vision tool. Then for each set, we aggregated the number of images, views, viewed datasets, and users based on three criteria: viewing date (date that the image was viewed), dataset date (date that the image was taken), and $D$ (difference in days). We now present three interesting findings.

First, while human-generated images were suitable for initiating community engagement, algorithm-generated images were useful for maintaining community engagement. In Figure \ref{fig:analysis-diff-time}, we aggregated number of views based on $D$, difference in days. The top graph in Figure \ref{fig:analysis-diff-time} showed that a large portion of views of human-generated images had small $D$, which indicated a short period between when a user viewed an image and when the image was taken. This suggested that our users tended to create animated images manually by using the thumbnail generator after a recent event (e.g. smoke emission), which showed the purpose of initiating community engagement. However, most of the views of algorithm-generated images had high $D$ (see the bottom graph in Figure \ref{fig:analysis-diff-time}). This showed that community members tended to use images generated automatically by the computer vision tool to review events occurring well beforehand, which demonstrated the objective of maintaining community engagement.

Second, the computer vision tool encouraged community members to explore more datasets. In Figure \ref{fig:analysis-dataset-date}, we aggregated the number of views based on dataset date, the time that the image was taken. The top and bottom graphs in Figure \ref{fig:analysis-dataset-date} show results for human-generated and algorithm-generated images respectively. By comparing these graphs, the number of views of algorithm-generated images were more distributed across datasets than the ones of human-generated images, which were concentrated on specific days.

\begin{figure}[t]
	\centering
	\includegraphics[width=1\columnwidth]{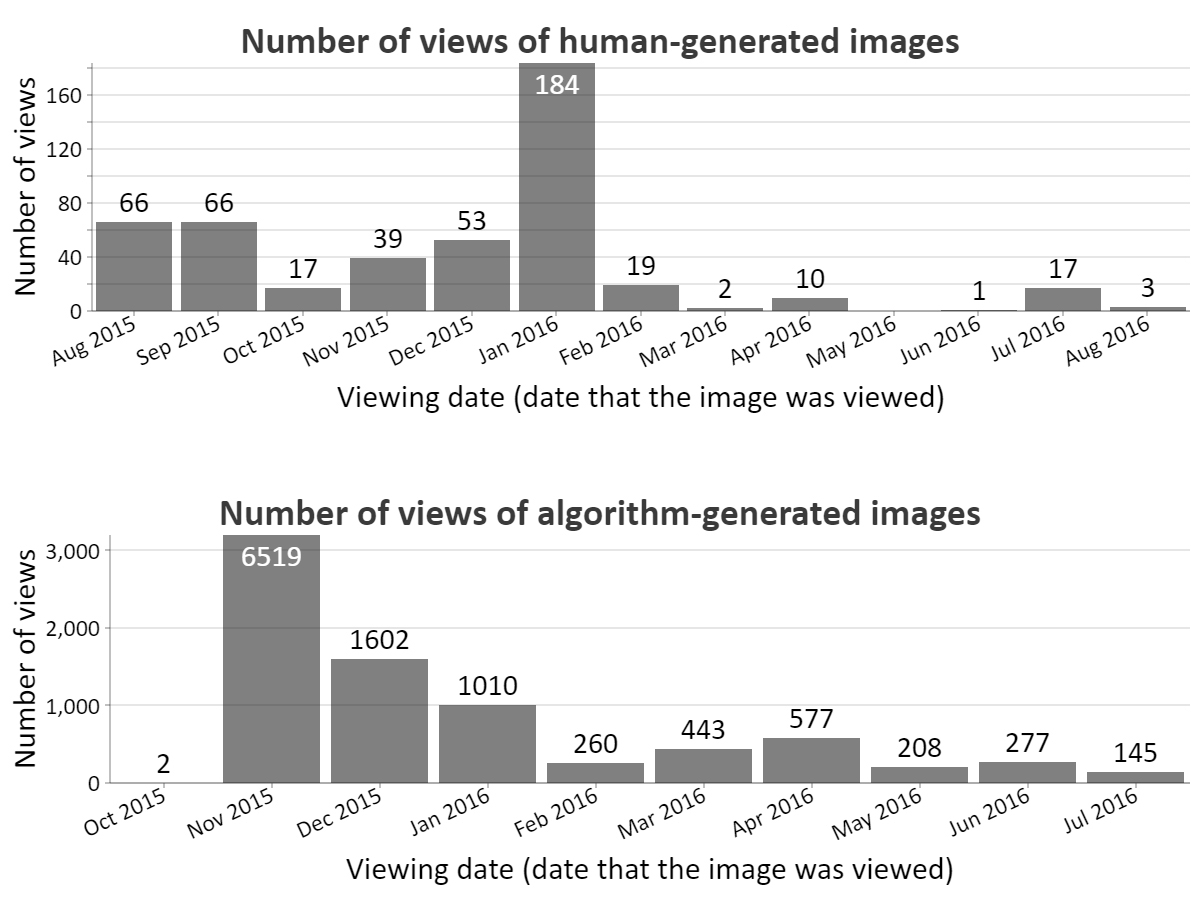}
	\caption{Number of views of human-generated or algorithm-generated images which are aggregated by viewing date. There is a significant decrease after January 2016, which was when the coke plant was closed.  }
	\label{fig:analysis-viewing-date}
\end{figure}

Third, the existence of the coke plant was significant in motivating the community to interact with the monitoring system. In Figure \ref{fig:analysis-viewing-date}, we aggregated the number of views based on viewing date, the time that image was viewed. The figure shows that community members viewed much less human-generated and algorithm-generated images after Jan 2016, which was the time that the coke plant was closed.

\subsubsection{User-based Sub-study}

For the user-based sub-study, we aggregated the number of images, views, and viewed datasets based on unique IP addresses to obtain a series of vectors. To find relationships, we computed the correlation matrix of five vectors into the number of: created human-generated images, viewed human-generated images, viewed datasets in human-generated images, viewed algorithm-generated images, and viewed datasets in algorithm-generated images. We now summarize two findings.

First, there were strong correlations within the usage of human-generated images. Community members who created more images by using the thumbnail generator also viewed more human-generated images (Pearson's R Correlation = 0.91) and explored more datasets (Pearson's R Correlation = 0.89). Moreover, community members who viewed more human-generated images also explored more datasets (Pearson's R Correlation = 0.8).

Second, it appeared that there was no obvious relationship between the usage of human-generated and algorithm-generated images. Community members who created or viewed more human-generated images did not necessarily view more algorithm-generated images (Pearson's R Correlation = 0.13 and 0.07 respectively). Furthermore, there were no strong correlations within the usage of algorithm-generated images. Community members who viewed more algorithm-generated images did not necessarily explore more datasets (Pearson's R Correlation = 0.35). The rhetorically compelling power of human-generated data should not be underestimated.

\subsection{Survey Study}

We now discuss the survey study for evaluating changes in the community's attitude after the intervention of our system.

\subsubsection{Participants}

ACCAN members were the primary users of the air quality monitoring system. Adult volunteers (age 18 and older) were recruited from these users through a Google Groups email. The email described the research purpose and included a link to an online survey. Paper surveys were also provided at a community meeting. All responses were kept confidential and there was no compensation. There was a brief consent script to review before taking the survey. We received 24 responses in total from 83 community members on the Google Groups (29\% response rate). One invalid response which contained inconsistent answers and five incomplete ones were discarded. Most of the participants had a high education level and were over the age of 35 (see Table~\ref{tb:demographic} for demographics).

\begin{table}[t]
	\begin{center}
		\begin{tabular}{|l|c|c|c|c|c|c|c|c|}
			\hline  & \!\!\!\! 18-24 \!\!\!\! & \!\!\!\! 25-34 \!\!\!\! & \!\!\!\! 35-44 \!\!\!\! & \!\!\!\! 45-54 \!\!\!\! & \!\!\!\! 55-64 \!\!\!\! & \!\!\!\! 64-74 \!\!\!\! & \!\!\!\! 75+ \!\!\!\! & \!\!\!\!\! Sum \!\!\!\!\!\\
			\hline \!\!\!\! No degree \!\!\! & 0 & 0 & 0 & 0 & 0 & 1 & 0 & \cellcolor{gray}1\\
			\hline \!\!\!\! Bachelor \!\!\! & 1 & 1 & 1 & 0 & 2 & 2 & 0 & \cellcolor{gray}7\\
			\hline \!\!\!\! Master \!\!\! & 0 & 0 & 2 & 2 & 2 & 3 & 0 & \cellcolor{gray}9\\
			\hline \!\!\!\! Doctor \!\!\! & 0 & 0 & 0 & 0 & 0 & 0 & 1 & \cellcolor{gray}1\\
			\hline \!\!\!\! Sum \!\!\! & \cellcolor{gray}1 & \cellcolor{gray}1 & \cellcolor{gray}3 & \cellcolor{gray}2 & \cellcolor{gray}4 & \cellcolor{gray}6 & \cellcolor{gray}1 & \cellcolor{gray}18\\
			\hline 
		\end{tabular} 
	\end{center}
	\caption{Age and education level for the participants of 18 valid survey responses. Participants have a high education level in general.}
	\label{tb:demographic}
\end{table}

\subsubsection{Procedure and Materials}

Participants filled out a survey. The survey was expected to take less than 30 minutes and contained three question types. The first type measured participants' involvement in the community action, such as exploring, documenting, and sharing data on the system. The second type measured community engagement, which included Likert scale questions related to the dependent variables: awareness, self-efficacy \cite{Bandura1977,Carroll2005}, and sense of community \cite{McMillan1986}. The third type asked demographics, such as age range and education level. The range of the Likert scale was from 1 to 5, with 5 being the highest attitude.

\subsubsection{Analysis}

\begin{figure}[t]
	\centering
	\includegraphics[width=1\columnwidth]{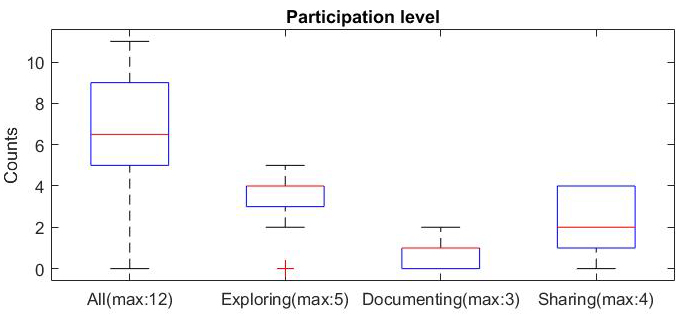}
	\caption{The boxplot of the participation level. We asked three multi-choice questions related to how users explore, document, and share the data provided by the system (the x-axis). These three questions had 5, 3, and 4 choices respectively. We summed up the number of choices that were selected by participants in each question to obtain participation levels (the y-axis). In general, the users had high participation levels.}
	\label{fig:participation}
\end{figure}

In the survey, participants answered three questions about how they explored, documented, or shared data by using the system. These three questions contained 5, 3, and 4 choices respectively. We summed up the number of choices that were selected by participants in each question to obtain participation levels (see Figure~\ref{fig:participation}). We also asked questions about the frequency (from 1 to 5, with 5 being the highest frequency) of browsing the data in the system after noticing bad smells, number of people that a participant discussed the system with, and number of monthly meetings (from 0 to 12) attended in 2015 (see Table.~\ref{tb:indep_var}).

\begin{table}[t]
	\begin{center}
		\begin{tabular}{|c|c|c|c|}
			\hline  & \!\!\! Browsing ($V_b$) \!\!\! & \!\!\! People discussed ($V_d$) \!\!\! & \!\!\! Meetings ($V_m$) \!\!\! \\
			\hline \!\!\! $\mu|\sigma$ \!\!\!\! & 2.94$|$1.35 & 22.28$|$21.85 & 7.83$|$3.60 \\
			\hline 
		\end{tabular}
	\end{center}
	\caption{The mean ($\mu$) and standard deviation ($\sigma$) of other independent variables. $V_b$ is the frequency (from 1 to 5, with 5 being the highest) of browsing the data in the system after noticing bad smells. $V_d$ is the number of people that a participant discussed the system with. $V_m$ is the number of monthly community meetings (from 0 to 12) attended in 2015. In general, participants were active in the community.}
	\label{tb:indep_var}
\end{table}

\begin{figure}[t]
	\centering
	\includegraphics[width=1\columnwidth]{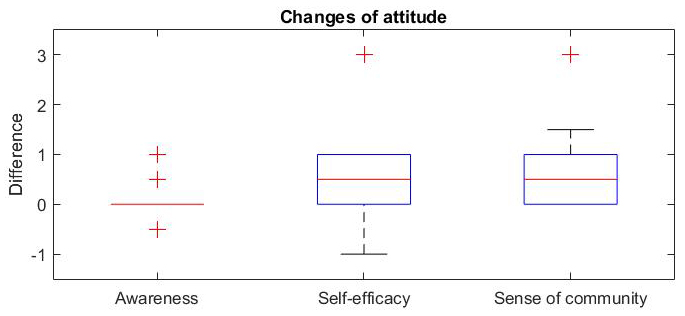}
	\caption{The boxplots of the changes of mental states among all participants after interacting with the monitoring system. The x-axis indicates dependent variables. The y-axis is the differences in Likert scale. Positive values mean increases, and vice versa.}
	\label{fig:attitude}
\end{figure}

For a dependent variable, participants answered a question set twice based on the time before (denote $S_i^b$) and after (denote $S_i^a$) they learned about the air quality monitoring system. Each question set had two Likert scale questions. We then averaged the Likert scales in set $S_i^b$ and $S_i^a$ to obtain a pair of scores. Figure~\ref{fig:attitude} showed the difference of scores for each dependent variable. Positive values indicated increases, and vice versa.


Our directional null hypotheses were that the community did not have significant increases in awareness, self-efficacy, and sense of community. Since the differences of our paired samples did not follow a normal distribution (see Figure \ref{fig:attitude}), we performed a right-tailed Wilcoxon signed-rank test, a nonparametric version of paired t-test. Table~\ref{tb:analysis} showed the p-values and confidence interval.

\begin{table}[t]
	\begin{center}
		\begin{tabular}{|l|c|c|c|}
			\hline  & Awareness & Self-efficacy & Community sense \\
			\hline p-value & 0.2500 & \cellcolor{gray}0.0042 & \cellcolor{gray}0.0010 \\
			\hline CI & 0.08$\mypm$0.15 & \cellcolor{gray}0.53$\mypm$0.40 & \cellcolor{gray}0.56$\mypm$0.38 \\
			\hline 
		\end{tabular} 
	\end{center}
	\caption{The p-value of right-tailed Wilcoxon signed-rank test and the confidence interval on the differences of paired samples. CI indicates 95\% confidence interval. Gray cells indicate statistical significance ($p<0.05$) or the confidence interval which does not contain zero.}
	\label{tb:analysis}
\end{table}

\subsubsection{Results}

According to the analysis (see Table~\ref{tb:analysis}), the result favored the alternative hypotheses, which claimed there were significant increases ($p<0.05$) in self-efficacy and sense of community after interacting with the system. The average increases in these two dependent variables were 0.53 and 0.56 respectively in Likert scale. However, we retained the null hypothesis, which stated there was no significant increase in awareness, since $p>0.05$ and the confidence interval contained zero.

Open-ended answers in surveys showed that the monitoring system could encourage agonistic discussion with regulators and empower the community in supporting local policy making. With the system, community members could report concrete scientific evidence of fugitive emissions to the local health department, such as animated smoke images and the exact time of emissions, instead of vague reports.

\begin{quote}
\textit{"I \highlight{made screenshots} of the [system name] dashboard at different times/days when wind was strong and in the direction of my community. I inserted these screenshots into Powerpoint slides. I \highlight{shared printed versions of these slides} with my Township commissioner when asking for assistance in reducing emissions."}

\textit{"I \highlight{continually spoke} at regional meetings, City, County, Health Department, Clairton, Lawrenceville, etc. Wrote numerous letters to the editor, most did get published, not all."}

\textit{"I reported specific emissions from [coke plant name] to ACHD. I was able to \highlight{provide specific times} so that ACHD could review the exact episodes that I was reporting."}

\textit{"I \highlight{shared web links} to the [system name] when I submitted complaints to the health department"}

\textit{"Confronted ACHD staffers repeatedly with \highlight{'uncomfortable' info}."}

\textit{"I \highlight{e-mailed images} to others, including regulators."}
\end{quote}

Moreover, others mentioned that their confidence in taking action was significantly improved after interacting with the system. One important reason was that integrating heterogeneous data (smoke images, air quality data, smell reports, and wind information) formed strong scientific evidence, which was powerful in communicating with regulators and thus changed the power relationship between citizens and the government.

\begin{quote}
\textit{"I felt that the more information/proof that I made available might help justify my concern and spur action. I \highlight{felt} that my concerns with what I was experiencing were \highlight{grounded in actual imagery, wind data and spatial data}."}

\textit{"I \highlight{believe} that the [system name] was very important in helping us \highlight{get the attention of regulators} (ACHD and EPA) and get them to take our concerns seriously."}

\textit{"The [system name] was one of the most important tools the community has in holding the plant accountable. I \highlight{believe} that images presented at the Nov. 2015 EPA ACHD ACCAN meeting \highlight{provided a tipping point for the plant's shutdown}."}

\textit{"I \highlight{believe} that the [system name] images shown at the November 2015 community meeting 'tipped the balance' for the EPA and may have resulted directly in the closing of [coke plant name]. In fact, \highlight{without those images, it may have taken years to close the plant}."}
\end{quote}

In addition, several community members specifically identified the political and educational values of the monitoring system. In addition, they showed a desire of reproducing the monitoring system on other neighborhoods.

\begin{quote}
\textit{"Background as a \highlight{environmental law paralegal}."}

\textit{"Fantastic \highlight{educational} tool."}

\textit{"I would like to see \highlight{similar monitoring of other pollution sites} in Pittsburgh, ie. the [other coke plant name] and others mentioned in the Toxic Ten listing."}
\end{quote}

\section{Discussion}

\begin{table}
	\begin{center}
		\begin{tabular}{|l|c|}
			\hline  & $\mu|\sigma$ \\
			\hline \!\!\! The timelapse video & \!\!\! 4.81$|$0.54 \!\!\! \\
			\hline \!\!\! Zooming in and out of the video & \!\!\! 4.50$|$0.73 \!\!\! \\
			\hline \!\!\! Sharing a web link of a view and time & \!\!\! 4.43$|$0.85 \!\!\! \\
			\hline \!\!\! Smell reports & \!\!\! 4.38$|$0.81 \!\!\! \\
			\hline \!\!\! Line charts showing sensor readings & \!\!\! 4.31$|$0.87 \!\!\! \\
			\hline \!\!\! The map showing sensor values & \!\!\! 4.44$|$0.73 \!\!\! \\
			\hline \!\!\! The thumbnail tool & \!\!\! 4.19$|$0.83 \!\!\! \\
			\hline \!\!\! The automatic smoke detection tool & \!\!\! 4.31$|$0.70 \!\!\! \\
			\hline \!\!\! Smoke images shown on the meeting with EPA \!\!\! & \!\!\! 4.94$|$0.25 \!\!\! \\
			\hline 
		\end{tabular} 
	\end{center}
	\caption{The mean and standard deviation ($\mu|\sigma$) of the importance rating of features on the air quality monitoring system. In general, participants rated all features important.}
	\label{tb:feature}
\end{table}

The community that we collaborated with has fought for decades to resolve the air pollution problem, which existed since 1999. The monitoring system was launched in Fall 2015. In November 2015, the community held a meeting at their local church with government officials from the ACHD (Allegheny County Health Department) and the EPA. During the meeting, as information technology supporters, we demonstrated the system and the visualization. In addition, the community projected hundreds of animated smoke images generated by the system on a large screen in front of ACHD and EPA regulators. Community members described how their living quality was affected by the air pollution together with animated smoke images, air quality sensors, crowdsourced smell reports, and wind data. The scientific knowlege demonstrated how heavy air pollution flowed into the neighborhood. The community successfully combined personal experiences and scientific evidence into a story to convince regulators. The story showed that the pollution source was the coke plant, and its fugitive emissions acturally affected the local air quality. This forced regulators to respond to the air quality problem publicly. The acting director of the EPA from the Region III Air Protection Division in Philadelphia pointed at the screen and said: \textit{"But what I see in the video, is totally unacceptable."} In addition, the local air quality problem became available for further debate and investigation. The administrator agreed that the EPA would continue to review the coke works' compliance with the 2012 federal consent decree. Furthermore, on December 2015, the parent company of the coke works announced the closure of the plant, which was the ultimate goal that the community had tried to achieve for decades.

\subsection{Insights}

Based on the major community meeting described in the previous paragraph and the results presented in the previous section, we now summarize our findings into three key insights and offer suggestions to future researchers.

\subsubsection{Use a Flexible and Iterative Design Process}
We encourage using a flexible and iterative procedure instead of a single and prescribed one. This practice is also mentioned by DiSalvo et al. \cite{DiSalvo-Louw-2009} as community co-design, a process which involves community members when designing a system that supports citizen empowerment. Often there are attempts to duplicate successful systems in another similar real-world context. However, this is unlikely to succeed because the environmental problem that the community deals with is wicked \cite{Rittel1973, Conklin-2005}. Every wicked problem has no clear formulation, is unique, and cannot be fully observed. Therefore, like the experience we describe in the design process and system sections, we recommend scheduling multiple design phases to reveal unique challenges and to apply specific solutions on these challenges iteratively. In the survey study, participants rate the importance of features of the system (see Table~\ref{tb:feature}). The rating scale is from 0 to 5, with 5 being the most important. The average ratings are all above 4, which verifies that the iterative design process help develop altogether useful system features to the community.

\subsubsection{Initiate and Maintain Community Engagement}
It is critical to initiate and maintain community engagement via actual participation in using the system. We recommend combining manual and automatic approaches, which are the thumbnail generator and the computer vision tool respectively in this work, to serve two different purposes in citizen participation. First, a manual approach can initiate citizen participation and lead to follow-up interactions. The image usage study shows that community members use the thumbnail generator to manually create images after they notice unusual events (see Figure \ref{fig:analysis-diff-time}), such as industrial smell or hazardous smoke. Correlation analysis of image usage indicates that users who create more images also view more images and explore more datasets (see the User-based Sub-study subsection). Second, an automatic approach can encourage community members to participate in a long temporal horizon. Smoke images generated automatically by the computer vision tool are used for reviewing fugitive emissions (see Figure \ref{fig:analysis-diff-time}). The computer vision tool encourages community members to explore more datasets (see Figure \ref{fig:analysis-dataset-date}). However, it appears that there are no clear correlations between the manual and automatic approach (see the User-based Sub-study subsection). How to integrate these two approaches seamlessly to open up and maintain citizen participation remains an important research question.

\subsubsection{Enable the Formation of Scientific Knowledge via Hybrid Data}
Data requires being interpreted into scientific knowledge to be impactful in changing unbalanced power relations between citizens and governments. Besides collecting data, providing affordance for citizens to make sense of the relationship among various types of data is key to generating scientific knowledge. We suggest integrating image, sensor, and crowdsourced data from both human and machines into such a system. Analysis in the survey study is limited by the small sample size of total users, and this should be taken as a caveat in regards to analysis of statistical significance. Nonetheless, Figure \ref{fig:attitude} shows the changes of participants' attitudes and Table \ref{tb:analysis} includes statistical significance findings in self-efficacy and sense of community. Open responses in the survey show that with scientific knowledge, citizens can present data in meaningful ways to regulators who have the power to make policy changes. At the meeting in November 2015, the community successfully influenced the attitude of the government after presenting the evidence. Scientific knowledge gives citizens power to advocate for their living quality and to influence other stakeholders.

\subsection{Limitation}

Measuring information and communication technology (ICT) interventions in community advocacy is generally challenging. Community advocacy has the ultimate goal of policy change, yet it is difficult to causally prove how critical to a successful policy change the communities' actions have been. Such projects succeed not only when policy goals are achieved, but in how the relationship between citizens, policy makers, and businesses evolves. This work shows that making scientific data transparent to stakeholders can foster sustainable relationships among them. It is sustainable in the sense that the system promotes a healthy and balanced power structure for democracy in the long term. We believe patterns of scientific data usage and changes of mental state among community members are useful proxies for evaluating the effectiveness of such projects. To better understand usage patterns, we suggest tracking the usage of data in the system. Future research about how to evaluate ICT interventions is still needed. For instance, qualitative research, like in-depth interviews, will be needed to identify key factors for successful collaboration between stakeholders and to understand changes of power dynamics among citizens, scientists, developers, and regulators. Moreover, forming scientific knowledge about the relationship between the smoke emissions and the severity of the air pollution by using the monitoring system currently relies on human interpretation. Additional future research involves enhancing the knowledge by analyzing the correlations between various types of data. The analysis can explain how these data reinforce or conflict with each other, which provides strong statistical scientific evidence.

Another limitation is that the sample size of participants in the survey study is too small and the statistical analysis conclusion (see subsection Results) is weak. Participants only represent a fraction of the population in the neighborhood near the coke works. They have high education (see Table~\ref{tb:demographic}) and involvement levels (see Table~\ref{tb:indep_var} and the left-most boxplot in Figure~\ref{fig:participation}), which includes interacting with the system, discussing the system with others, and attending monthly community meetings. Most of them have strong activation before learning about the monitoring system, which causes the failure to reject the null hypothesis related to awareness (see Table \ref{tb:analysis}). The strong activation may also result in the high correlation between community members who created and viewed smoke images (see subsection User-based Sub-study). Nevertheless, one alternative explanation of this limitation is that without high awareness, it would be impossible to support community advocacy with ICT interventions. In other words, high awareness may be a necessary condition for successful citizen empowerment. How attitude may change among people with low education or low involvement level after interacting with the air quality monitoring system still remain an open research question.

Furthermore, the smoke detection algorithm used in the system is tuned to operate in our settings. Currently, the algorithm uses a heuristic method and has too many tuning parameters, which is not robust enough for similar contexts for other communities. One approach to generalize the system is to collect crowdsourced labels via mobile or online platforms, which requires deeper citizen participation. These labels can then be used to train a smoke image classifier using machine learning. Moreover, it appears that the existence of the coke plant is the major source of motivation (see Figure \ref{fig:analysis-viewing-date}). This crowdsourcing approach may provide extra motivations to the community. Besides collecting labels, organizing the hybrid scientific data collected in the system into a comprehensive dataset can potentially assist future academic research related to environmental problems.

\section{Conclusion}

This paper presents a web-based air quality monitoring system which integrates image, sensor, and crowdsourced data. It is an instance of adversarial design \cite{DiSalvo2010, DiSalvo2012} which critically reveals, questions, and challenges a real-world environmental problem. The system provides technological affordance for forming strong scientific evidence. We discuss the iterative participatory design process that leads to decisions of system features with the community. We describe our evaluation, which includes an image usage study from server logs and a survey study. The survey study indicates statistically significant increases in self-efficacy and sense of community among users after interacting with the system. Open responses in the study show that the system promotes critical discussions with policy makers and empowers citizens to participate in community actions. Based on the evaluation, we offer three key insights about using an iterative design process, encouraging community engagement, and forming scientific knowledge. Finally, we mention limitations and future research directions related to evaluating the intervention of information technology, studying user behavior of community members with low participation level, and generalizing the smoke detection algorithm by collecting crowdsourced labels. We hope that this work can inspire other researchers to contribute towards developing innovative information technology that supports citizen empowerment.

\section{Acknowledgments}

The Heinz Endowments, Allegheny County Clean Air Now, and all other participants. The authors thank Yen-Chi Chen for the advice in statistical analysis.

%
%
%
%
%
\balance{}

\bibliographystyle{SIGCHI-Reference-Format}
\bibliography{ref}

\end{document}